\documentclass{article}

\usepackage{PRIMEarxiv}

\usepackage[utf8]{inputenc} 
\usepackage[T1]{fontenc}    
\usepackage{hyperref}       
\usepackage{url}            
\usepackage{booktabs}       
\usepackage{amsfonts}       
\usepackage{nicefrac}       
\usepackage{microtype}      
\usepackage{lipsum}
\usepackage{fancyhdr}       
\usepackage{graphicx}       
\graphicspath{{media/}}     

\usepackage{tikz}
\usetikzlibrary{shapes.geometric,shapes.arrows,decorations.pathmorphing}
\usetikzlibrary{matrix,chains,scopes,positioning,arrows,fit}

\usepackage{ucs}
\usepackage{amsmath}
\usepackage{amsfonts}
\usepackage{amssymb}
\usepackage{enumitem}
\usepackage[cal=boondox]{mathalfa} 
\usepackage{bm}
\newcommand*{\B}[1]{\ifmmode\bm{#1}\else\textbf{#1}\fi}

\usepackage{mathtools}

\usepackage{amsthm}

\newtheorem{proposition}{Proposition}
\newtheorem{remark}{Remark}

\newtheorem{theorem}{Theorem}

\title{On the Metric Properties of IR Evaluation Measures Based on Ranking Axioms
}

\author{
  F. Giner \\
  E.T.S.I. Informática UNED, \\
  C/ Juan del Rosal, 16, 28040 - Madrid, Spain \\
  \texttt{fginer3@alumno.uned.es}   
}

\tikzstyle myBG=[line width=3pt,opacity=1.0]

\newcommand{\precdot}{\prec\mathrel{\mkern-5mu}\mathrel{\cdot}}

\makeatletter
\newcommand{\preceqdot}{\mathrel{\mathpalette\pr@ceqd@t\relax}}
\newcommand{\pr@ceqd@t}[2]{%
  \begingroup
  \sbox\z@{$#1\prec$}\sbox\tw@{$#1\preceq$}%
  \dimen@=\dimexpr\ht\tw@-\ht\z@\relax
  {\preceq}%
  \mkern-5mu
  \raisebox{\dimen@}{$\m@th#1\cdot$}%
  \endgroup
}
\makeatother

\begin{document}
\maketitle

\begin{abstract}
The axiomatic analysis of IR evaluation metrics has contributed to a better understanding of their properties. Some works have modelled the effectiveness of retrieval measures with axioms that capture desirable properties on the set of ranked lists of documents. Recently, in \cite{giner2022effect}, it has been shown that three of these axioms lead to some orderings. This work formally explores the metric properties of the set of rankings, endowed with these orderings. Based on lattice theory, the possible metrics and pseudo-metrics, defined on these structures, are determined. It is found that, when the relevant documents are prioritized, precision, recall, $RBP$ and $DCG$ are metrics on the set of rankings, however they are pseudo-metrics when the swapping of documents is considered.
\end{abstract}


\section{Introduction}
In the Cranfield paradigm \cite{cleverdon1967cranfield}, relevance judgements are assigned to a ranked list of documents for a topic. Then, an IR evaluation metric can be seen as a mapping that relates the set of possible lists of judged documents (empirical domain) with the real numbers (scores). They quantify, in numeric terms, the ability of systems to leave aside the irrelevant documents from the relevant ones.

There have been proposed a very large number of measures which evaluate different aspects of IR systems \cite{demartini2006classification,demartinisurvey,amigo2014general,amigo2009comparison}. In designing these measures, much of the effort has gone into capture the \emph{effectiveness} of retrieval systems. Sometimes, these attempts have led to pay less attention to the three postulates of a metric\footnote{Here, the mathematical term of ``metric'' collides with the commonly used term of IR evaluation ``metric''. To solve this issue, in this paper, the term ``metric'' will have its mathematical sense. The term ``measure'' will design any mapping, from the set of rankings to the real numbers. Thus, IR evaluation metrics will be referred as IR evaluation measures.}, in the mathematical sense: \emph{identity of indiscernible}, \emph{symmetry} and \emph{triangular inequality}. This can have inadequate consequences, such as drawn counter-intuitive conclusions.

The metric properties of a mapping are directly related to the properties of the set where it is defined \cite{frechet1906quelques, hausdorff2005set}: \emph{if there exists a mapping or function that verifies the three postulates, then the set where it is defined is a} metric space \emph{and the function that verifies these properties is a} metric. The importance of the empirical domain can also be observed on the IR evaluation measures, for example, in the statement ``an evaluation measure is an ordinal scale'', the evaluation measure is preserving the \emph{existing ordering} between rankings of documents.

Some works \cite{bollmann1984two,moffat2013seven,amigo2013general,ferrante2015towards} have presented desirable properties that any reasonable IR evaluation measure should satisfy, for instance, ``it is preferred to retrieve relevant documents in the top ranking positions''. These properties (axioms, constraints or heuristics) lead to different orderings in the set of ranked lists of documents \cite{giner2022effect}. Then, some obvious questions arise: Is the set of rankings, endowed with this orderings, a metric space? Which are the corresponding metrics? Are the values assigned by the IR evaluation measures \emph{consistent} with the orderings derived from the desirable properties? Here, the \emph{consistency} can be understood as meaning that the scores assigned by the IR evaluation measures do not contradict the orderings derived from the desirable properties. 

This paper determines if the set of ranked lists of documents, endowed with the orderings derived in \cite{giner2022effect}, is a metric or pseudo-metric space, i.e., if there exists at least one metric or pseudo-metric defined on it. Then, in the set of rankings where it is feasible, all the possible metrics\slash pseudo-metrics are inferred. To this end, it is presented the concept of valuation, which is the ground of the concepts of \emph{measure} and \emph{probability} in \emph{measurement theory}. There are shown explicit expressions of the mappings, in terms of a remarkable subset of rankings presented in \cite{giner2022effect}: the join-irreducible elements. Then, it is checked if some IR evaluation measures are metrics or pseudo-metrics. It is found that precision, recall, $RBP$ and $DCG$ are metrics on the set of rankings where the relevant documents are prioritized, however they are pseudo-metrics when the swapping of documents is considered.

The rest of the paper is organized as follows: In Section \ref{sec:SoA}, some related works are reported. In Section \ref{sec:lattice}, we briefly introduce notations and some basic results of lattice theory. In Section \ref{sec:notation}, the problem is contextualized in the IR setting. Then, in Sections \ref{sec:set-based} and \ref{sec:rank-based}, it is determined when precision, recall, $RBP$ or $DCG$ are metrics for the set-based and rank-based retrieval respectively. Finally, in Section \ref{sec:conclusion} some conclusions are drawn.

\section{Lattice Theory}
\label{sec:lattice}
In this section we recall some basic notions of lattice theory, for additional background, we refer the reader to the  textbooks \cite{birkhoff1940lattice, gratzer2002general}. 

A \emph{partially ordered set} or \emph{poset} is a non-empty set, $X$, together with a binary, reflexive, antisymmetric and transitive relation, $\preceq$, defined on $X$. A poset $(X, \preceq)$ is called a \emph{totally ordered set} or a \emph{chain} if any two elements, $x$, $y \in X$, are comparable, i.e., $x \preceq y$ or $y \succeq x$.

A (closed) interval is a subset of $X$ defined as $[x, y] = \{z \in X : x \preceq z \preceq y\}$.
It is said that $y$ \emph{covers} $x$ in $X$, denoted by $x \precdot y$, if there does not exist $z \in X$ such that $x \neq z \neq y$ and $x \preceq z \preceq y$. An interval $[x, y]$ is called \emph{prime}, if $x \precdot y$, i.e., $[x, y] = \{x, y\}$. 

Thus, $(X, \preceq)$ has associated a graph, called its \emph{Hasse diagram}, where nodes are labelled with the elements of $X$ and the edges indicate the covering relation. Nodes are represented in different levels, if $y$ covers $x$, then $x$ is below $y$ in the diagram. A poset $X$ is called \emph{graded} if there exists an integer-valued function, $\rho$, defined on $X$, such that, if $x \precdot y$, then $\rho(y) = \rho(x) + 1$. The function $\rho$ is called the \emph{rank} function.

An upper bound for a subset, $S \subseteq X$, is an element, $x \in X$, for which $s \preceq x$, $\forall s \in S$. A lower bound for a subset of a poset is defined analogously. The \emph{greatest lower bound} or \emph{meet} of two elements $x, y \in X$ is denoted $x \wedge y$. The \emph{least upper bound} or \emph{join} of two elements $x, y \in X$ is denoted $x \vee y$. It is said that $X$ has a \emph{top element}, denoted by $1 \in X$, if $x \preceq 1$, $\forall x \in X$; dually, $0 \in X$ is the \emph{bottom element}, if $0 \preceq x$, $\forall x \in X$.

Considering the join and the meet as operators, it is said that $(X, \wedge, \vee, \preceq)$ is a \emph{lattice} if $(X, \preceq)$ is a poset and every two elements have a meet and a join. A lattice is \emph{modular} if $x \preceq y$ implies $x \vee (z \wedge y) = (x \vee z) \wedge y$, where $x$, $y$ and $z$ are arbitrary elements. A lattice is \emph{distributive} if its meet operator distributes over its join operator, i.e., $x \wedge (y \vee z)$ $=$ $(x \wedge y) \vee (x \wedge y)$, $\forall x$, $y$, $z \in X$. The distributive property is a stronger condition than the modular property; thus, any distributive lattice is a modular lattice \cite{birkhoff1940lattice}. An example of distributive lattices are the chains \cite{gratzer2002general}. 

There are some remarkable elements in a lattice, given $j \in X$ (with $j \neq 0$), it is said to be a \emph{join-irreducible} element, if $x \vee y = j$ implies $x = j$ or $y = j$. Thus, $j \in X$ is join-irreducible if it cannot be expressed as the join of two elements that are strictly smaller than $j$. The set of the join-irreducible elements of $X$ is denoted by $J_{X}$\footnote{The join-irreducible elements are those that have only one descendant in the Hasse diagram of $X$ \cite{gratzer2002general}}. Let $x$ be an element of $X$, it could be useful to consider the subset of join-irreducible elements smaller than $x$, which will be denoted by $J_{x} = \{j \in J_{X} : j \preceq x\}$. The following result highlights the importance of the join-irreducible elements of a finite distributive lattice.

\begin{theorem}\label{th:join-join-irreducible}.S. Blyth \cite{blyth2005distributive}]
If $X$ is a finite distributive lattice, then every element of $X \setminus \{0\}$ can be expressed uniquely as an irredundant join of join-irreducible elements.
\end{theorem}

A real-valued mapping defined on a lattice, $v: X \rightarrow \mathbb{R}$, is a \emph{valuation}, if it satisfies the following property: $v(x) + v(y) = v(x \vee y) + v(x \wedge y) $, $\forall x$, $y \in X$. A valuation, $v: X \rightarrow \mathbb{R}$ is called \emph{isotone} or \emph{monotone} (resp. \emph{positive}), if $x \preceq y \Rightarrow v(x) \leq v(y)$, where $\leq$ denotes the ordering on $\mathbb{R}$ (resp. $x \preceq y$ and $x \neq y$ $\Rightarrow v(x) < v(y)$). 

The next result shows a characterization of the valuations in terms of the join-irreducible elements.
\begin{theorem}[G.C. Rota \cite{rota1971combinatorics}]
A valuation, defined on a finite distributive lattice, $X$, is uniquely determined by the values it takes on the set of join-irreducible elements of $X$, and these values can be arbitrarily assigned.
\end{theorem}

In \cite{caspard2012finite}, is given an explicit expression of these functions. Considering an assignment or mapping from the set of join-irreducible elements to the positive real numbers, $w:J_{X} \longrightarrow \mathbb{R}^{+}$, any isotone valuation, $v$, on a finite distributive lattice can be obtained as:
\begin{equation}
\label{eq:valuation}
v(x) = \sum_{j \in J_{x}} w(j)
\end{equation}

\begin{theorem}[G. Birkhoff \cite{birkhoff1940lattice}]
\label{th:pseudo-metric}
In any lattice, $L$, with an isotone valuation, the distance function, $d_v (x,y) = v(x \vee y) - v(x \wedge y)$, satisfies for all $x$, $y$ and $z \in X$: (i) $d_v(x,x) = 0$, $d_v(x, y) \geq 0$ and $d_v(x,y) = d_v(y,x)$; (ii) $d_v(x,z) \leq d_v(x, y) + d_v(y,z)$; and (iii) $d_v(z \vee x, z\vee y) + d_v(z \wedge x, z \wedge y) \leq d_v(x,y)$.
\end{theorem}

This result enables to define a \emph{pseudo-metric} or \emph{semi-metric} lattice as a lattice with an isotone valuation. A pseudo-metric lattice is so named because the distance, $d_v(x, y)$, verifies the properties of a pseudo-metric: symmetry, $d_v(x,y) = d_v(y, x)$, and triangular inequality, $d_v(x,y) \leq d_v(x,z) + d_v(z,y)$.

In addition, if $d_v(x,y)=0 \Leftrightarrow x=y$, then $d_v$ verifies the three postulates of a metric and $(X, \wedge, \vee, \preceq)$ is called a \emph{metric} lattice. As $v$ is an isotone valuation, this condition is equivalent to $x \wedge y \preceq x \vee y \Rightarrow v(x \wedge y) < v(x \vee y)$ \cite{birkhoff1940lattice}. Hence, a pseudo-metric lattice yields a metric lattice if and only if the valuation, which defines, is positive. 

\begin{theorem}[G. Birkhoff \cite{birkhoff1940lattice}]
\label{th:modular-metric}
Any metric lattice is modular.
\end{theorem}
The converse of this theorem holds for finite modular lattices \cite{rutherford1967introduction}. From this result, any finite distributive lattice is a metric lattice since distributive implies modular. Indeed, every isotone valuation, in a finite distributive lattice, is positive since, by Theorem \ref{th:join-join-irreducible} and Equation \ref{eq:valuation}, if $x \preceq y$ and $x\neq y$, then $J_x \subset J_y$. 

Let $v$ be any isotone valuation on a finite distributive lattice generated by a strictly positive real-assignment, $w$, of the join-irreducible elements, when each edge, from $x$ to $y$, of the covering graph is weighted by $\big\vert v(y) - v(x) \big\vert$, then the minimum path length metric coincides with $d_v$ and it verifies:
\begin{equation}
\label{eq:metric}
d_{v}(x,y) = v(x) + v(y) - 2 \cdot v(x \wedge y) = \sum_{j \in J_{x} \triangle J_{y}} w(j)
\end{equation}
where $\triangle$ is the symmetric difference operator \cite{leclerc1994medians,leclerc1993lattice}. 

In addition, if $w$ is defined by $w(j) = k$ for all $j \in J_X$, $k \in \mathbb{R}^{+}$ and $X$ is a distributive lattice, then $\big\vert v(x) - v(y) \big\vert = k$ for every edge in the Hasse diagram of $X$. Thus, the valuation can be expressed as $v(x) = k \cdot \big\vert J_x \big\vert$ and $d_v$ is the natural distance on the covering graph, $d_{v}(x, y) = k \cdot \big\vert J_{x} \triangle J_{y} \big\vert$.

Finally, fixing one of the arguments to the bottom element, $0$, and choosing $k=1$, this valuation coincides with the rank function: $\rho(x) = d_v(0,x) = \big\vert J_x \big\vert$.

\section{IR Setting}
\label{sec:notation}
In this section, some notation needed throughout the paper is introduced. The formalism of \cite{ferrante2018general} is adopted.

Consider a finite set of documents that are retrieved from query terms provided by a user. An (ordered or not) collection of $N$ retrieved documents will be called a \emph{system run} of length $N$ for a topic.

Once documents have been retrieved, they can be classified by their relevance degree. Let  $(REL, \preccurlyeq)$ be a finite and totally ordered set of relevance degrees, where $REL = \{\mathcal{a}_0, \ldots, \mathcal{a}_{c}\}$ with $\mathcal{a}_{i} \prec \mathcal{a}_{i+1}$, for each $i \in \{1, \ldots, c-1\}$. These relevance degrees can be categorical labels. To handle numerical values, a \emph{gain function}, $g: REL \rightarrow \mathbb{R}^{+}$, is considered as a map that assigns a positive real number to any relevance degree. In this paper, it can be assumed that $g(\mathcal{a}_0) = 0$ and $g$ is a strictly increasing function. A particular case of such a map is the \emph{indicator function}, defined as $\delta_{\mathcal{a}}(\mathcal{a}_j) = j$, for each $j = 0, \ldots c$. 

When every retrieved document of a system run is classified with a relevance degree, a \emph{judged run} is obtained, denoted by $\hat{r}$. 

In the \emph{set-based retrieval}, each judged run is an unordered set of $N$ relevance degrees, which may contain the same element several times. For instance, for $N=4$ and $3$ relevance degrees, a judged run is given by $\hat{r} = \{\mathcal{a}_2, \mathcal{a}_1, \mathcal{a}_1, \mathcal{a}_0\}$. For the sake of clarity, the convention is used to represent $\hat{r}$ as $\{\hat{r}_1, \ldots, \hat{r}_N\}$, where $\hat{r}_i \succcurlyeq \hat{r}_{i + 1}$, for any $i \in \{1, \ldots N-1\}$, i.e., the relevance degrees are listed in decreasing order. This process does not affect the results obtained. The $j$-th element of the set $\hat{r}$ is denoted by $\hat{r}_j$.

In the \emph{rank-based retrieval}, each judged run is an ordered list of $N$ relevance degrees. For instance, for $N=4$ and $3$ relevance degrees, a judged run is given by $\hat{r} = (\mathcal{a}_1, \mathcal{a}_0, \mathcal{a}_2, \mathcal{a}_1)$. In this case, the $j$-th element of $\hat{r}$ is denoted by $\hat{r}[j]$.
 
The set of all the possible judged runs of length $N$ is a finite set, which will be denoted by $R(N)$.

In \cite{giner2022effect}, three commonly accepted operations about the set where an IR evaluation measure is defined are stated:
\begin{itemize}
\item \textbf{Replacement}: ``Replacing a document for another one with a higher relevance degree is a preferred option'' \cite{ferrante2015towards, bollmann1984two, amigo2018axiomatic}.
\item \textbf{Projection}: ``Removing all documents that are not considered in the highest or more relevant part of a list is a preferred option'' \cite{amigo2013general,amigo2018axiomatic}.
\item \textbf{Swapping}: ``Swapping a less relevant document in a higher rank position with a more relevant one in a lower rank position is a preferred option'' \cite{moffat2013seven, amigo2013general, ferrante2015towards}.
\end{itemize}

Combining these properties, five orderings are obtained: two for the set-based retrieval (\textbf{Projection+Replacement} and \textbf{Replacement}) and three for the rank-based retrieval (\textbf{Projection+Replacement}, \textbf{Replacement} and \textbf{Swapping+Replacement}).

$R(N)$ endowed with any of these orderings is a poset and considering the meet and join operations, $(R(N), \wedge, \vee, \preceq)$ can be considered as a lattice \cite{giner2022effect}. 

An IR evaluation measure can be seen as a mapping from $R(N)$ to the set of real numbers. In the set based retrieval some remarkable evaluation measures are as follows.
\begin{itemize}
\item Generalized Precision ($gP$) \cite{ferrante2018general}:
$$gP(\hat{r}) = \frac{1}{N} \cdot \sum_{i=1}^{N} \frac{g(\hat{r}_i)}{g(\mathcal{a}_{c})} \ .$$
This measure represent a user model which always reads from rank $1$ down to $k$, and then stops. This score represents the average rate at which a user accures relevance.
\item Generalized Recall ($gR$) \cite{ferrante2018general}
$$gR(\hat{r}) = \frac{1}{RB} \cdot \sum_{i=1}^{N} \frac{g(\hat{r}_i)}{g(\mathcal{a}_{c})} \ ,$$
where $RB$ is the recall base (total number of relevant documents). The user model of this measure is similar to $gP$, but the score represents the average rate that a relevant document will be retrieved. 
\end{itemize}
These evaluation measures can also be considered in the rank-based retrieval, in addition to the following. 
\begin{itemize}
\item Graded Rank-Biased Precision ($gRBP$) \cite{moffat2008rank,sakai2008information, ferrante2018general}:
$$gRBP(\hat{r}) = \frac{(1-p)}{g(\mathcal{a}_c)} \cdot \sum_{i=1}^N p^{i-1} \cdot g(\hat{r}[i]) \ ,$$
where $p$ reflects the persistence scanning a system output run. This score is numerically equal to the expected relevance of the last document inspected.

\item Discounted Cummulated Gain ($DCG$) \cite{kekalainen2002using}:
$$DCG_{b}(\hat{r}) = \sum_{i=1}^N \frac{g(\hat{r}[i])}{\max \{1, \log_{b}i \}} \ ,$$
where the parameter $b$ typically ranges over $2$ and $10$. The user is assumed to inspect an unbounded number of items. It models a user that scans down a ranked list, growing less interested with each successive rank; the gain function models the utility the user derives from each document.
\end{itemize}

The valuations on $(R(N), \wedge, \vee, \preceq)$ are real-valued mappings, $M: R(N) \longrightarrow \mathbb{R}$, satisfying $M(\hat{r} \vee \hat{s}) = M(\hat{r}) + M(\hat{s}) - M(\hat{r} \wedge \hat{s})$, $\forall \hat{r}$, $\hat{s} \in R(N)$. They are \emph{isotone} or \emph{monotone} (resp. \emph{positive}), if $\hat{r} \preceq \hat{s} \Rightarrow M(\hat{r}) \leq M(\hat{s})$, where $\leq$ denotes the ordering on $\mathbb{R}$ (resp. $\hat{r} \preceq \hat{s}$ and $\hat{r} \neq \hat{s}$ $\Rightarrow M(\hat{r}) < M(\hat{s})$).

\begin{remark}
An IR evaluation measure, $M:R(N)\longrightarrow \mathbb{R}$, is an \emph{ordinal scale} or \emph{order preserving}, if $\hat{r} \preceq \hat{s} \Leftrightarrow M(\hat{r}) \leq M(\hat{s})$. The difference with the definition of an isotone mapping is the implication from right to left, thus not every isotone mapping is an ordinal scale, however every ordinal scale is an isotone mapping. One of the aims of this paper is to determine all the isotone valuations (i.e., pseudo-metrics) defined on $R(N)$, thus the obtained results generalize the ordinal scales or even higher order scale types, which are valuations.
\end{remark}

To classify the metric properties of the empirical domain of IR evaluation measures, the following results of Section \ref{sec:lattice} are considered: (i) $(R(N), \wedge, \vee, \preceq)$ is a pseudo-metric lattice if and only if there exists an isotone valuation defined on $R(N)$; and (ii) $(R(N), \wedge, \vee, \preceq)$ is a metric lattice if and only if there exists a positive valuation defined on $R(N)$.

If $(R(N), \wedge, \vee, \preceq)$ is a finite distributive lattice and $w:J_{R(N)} \longrightarrow \mathbb{R}^{+}$ is an assignment to its join-irreducible elements, then any isotone valuation is as follows (see Section \ref{sec:lattice}).
\begin{equation}
\label{eq:valuation2}
M(\hat{r}) = \sum_{\hat{j} \in J_{\hat{r}}} w(\hat{j})
\end{equation}

These isotone valuations, in a finite distributive lattice, are really positive valuations; thus, they generate a metric, $d_M$, which coincides with the minimum path metric on the covering graph and is given by $d_{M}(\hat{r}, \hat{s}) = M(\hat{r}) + M(\hat{s}) - 2 \cdot M(\hat{r} \wedge \hat{s})$ (see Section \ref{sec:lattice}).

If $w$ is a constant assignment, with value $k \in \mathbb{R}^{+}$, then $d_{M}(\hat{r}, \hat{s}) = k \cdot \big\vert J_{\hat{r}} \triangle J_{\hat{s}} \big\vert$ is the natural distance on the Hasse diagram. In the particular case of $k=1$, this valuation is the rank function and it is given by $\rho(\hat{r}) = \big\vert J_{\hat{r}} \big\vert$.

In the following sections, it will be considered the five orderings derived in \cite{giner2022effect}. The key point is to determine if the empirical domain is a metric or pseudo-metric lattice. Then, the possible mappings on this set can be analysed. 

If the set of rankings, endowed with one of these orderings, is not a metric lattice, then there can be no positive valuations. However, it may exists isotone valuations. If there exists a pseudo-metric, then the set of rankings would be a pseudo-metric lattice. 

If the set of rankings, endowed with one of these orderings, is a metric space, then there is at least one mapping that is a positive valuation, which could be an ordinal scale. In addition, if the join-reducible assignment is constant, then it represents the natural distance. However, in a metric lattice there may also exist other valuations that need not be isotone or positive. Some IR evaluations measures may exhibit this situation, for instance, in a metric lattice, it could be an IR evaluation measure which is an isotone valuation, but not positive, i.e., it would be a pseudo-metric. 

\section{Set-based Retrieval}
\label{sec:set-based}
In this Section, the set-based retrieval is considered and the two orderings which were deduced in \cite{giner2022effect} are analysed.

\subsection{Projection+Replacement [set-based]}
Let $\hat{r}, \hat{s} \in R(N)$ such that $\hat{r} \neq \hat{s}$, and let $k$ be the largest relevance degree at which the two runs differ for the first time, i.e., 
$k = max \big\{j \leq c : \big\vert \{i : \hat{r}_i = \mathcal{a}_j\}\big\vert \neq \big\vert \{i : \hat{s}_i = \mathcal{a}_j\} \big\vert \big\}$; then, they are ordered by: 
$$\hat{r} \preceq \hat{s} \Longleftrightarrow \big\vert \{i : \hat{r}_i = \mathcal{a}_k\}\big\vert \leq \big\vert \{i : \hat{s}_i = \mathcal{a}_k\}\big\vert$$
A system run is preferred to another if for the highest relevance degree at which the two system runs differ, the first has more documents of this relevance degree. 

As $(R(N), \preceq)$ is a totally ordered set \cite{giner2022effect}, then every element is join-irreducible, except $\hat{0}$, i.e., $J_{R(N)} = R(N) \setminus \{\hat{0}\}$. Thus, $J_{\hat{r}}$ is the set of all the elements smaller than $\hat{r}$, except $\hat{0}$. 

$(R(N), \wedge, \vee, \preceq)$ is a metric lattice since it is a chain (see Section \ref{sec:lattice}). In the following, it will be determined all the possible isotone\footnote{They are really positive since $R(N)$ is a finite distributive lattice.} valuations on $R(N)$. Consider an assignment, $w:R(N) \setminus \{\hat{0}\} \longrightarrow \mathbb{R}^{+}$ and a judged run, $\hat{r} \in R(N) \setminus \{\hat{0}\}$, it can be assumed that $\hat{0} =\hat{r}^{0} \preceq \hat{r}^{1} \preceq \ldots \preceq \hat{r}^{k} = \hat{r}$. As every totally ordered set is a distributive lattice, then, by Equation \ref{eq:valuation2}, any isotone valuation can be obtained as:
$$M(\hat{r}) = \sum_{i = 1}^{k} w(\hat{r}^{i})$$

Considering $\hat{r}$, $\hat{s} \in R(N)$, each of these isotone valuations, $M$, generates a metric, $d_M$, which coincides with the minimum path metric and is given by $d_{M}(\hat{r}, \hat{s}) = M(\hat{r}) + M(\hat{s}) - 2 \cdot M(\hat{r} \wedge \hat{s})$. As $R(N)$, endowed with the \textbf{projection+replacement} [set-based] ordering, is a totally ordered set; then, it can be assumed that $\hat{r} \preceq \hat{s}$, i.e., $\hat{r} \wedge \hat{s} = \hat{r}$. Therefore, $d_{M}(\hat{r}, \hat{s}) = M(\hat{s}) - M(\hat{r})$.

If $w(\hat{r}) = k$, $\forall \hat{r} \in R(N) \setminus \{\hat{0}\}$, with $k \in \mathbb{R}^{+}$, then these valuations are expressed as $M(\hat{r}) = k \cdot \big\vert J_{\hat{r}} \big\vert$ and the natural distance on the covering graph is $d_{M}(\hat{r}, \hat{s}) = k \cdot \big\vert J_{\hat{s}} \triangle J_{\hat{r}} \big\vert$.

Thus, the isotone valuations, $M(\hat{r})$, with constant assignment to the join-irreducible elements count the number of elements from $\hat{0}$ to $\hat{r}$. It can be given an alternative expression of $M(\hat{r})$. Considering that $M(\hat{0}) = 0$, the \textbf{Projection+Replacement} [set-based] ordering is the following: $\{\mathcal{a}_0, \mathcal{a}_0, \mathcal{a}_0, \ldots, \mathcal{a}_0\} \precdot \{\mathcal{a}_1, \mathcal{a}_0, \mathcal{a}_0, \ldots, \mathcal{a}_0\} \precdot \{\mathcal{a}_1, \mathcal{a}_1, \mathcal{a}_0, \ldots, \mathcal{a}_0\} \precdot \cdots \precdot \{\mathcal{a}_1, \mathcal{a}_1, \mathcal{a}_1, \ldots, \mathcal{a}_1\} \precdot \{\mathcal{a}_2, \mathcal{a}_0, \mathcal{a}_0, \ldots, \mathcal{a}_0\} \precdot \{\mathcal{a}_2, \mathcal{a}_1, \mathcal{a}_0, \ldots, \mathcal{a}_0\} \precdot \{\mathcal{a}_2, \mathcal{a}_1, \mathcal{a}_1, \ldots, \mathcal{a}_0\} \precdot \cdots$. For example, if $\hat{r}= \{\mathcal{a}_k, \mathcal{a}_0, \ldots, \mathcal{a}_0\}$, the number of elements from $\hat{0}$ to $\hat{r}$ are the combinations of $N$ elements taken $\delta_{\mathcal{a}}(\mathcal{a}_k)= k$ at a time, with repetitions, i.e., $\binom{\delta_{\mathcal{a}}(\hat{r}_0) + N - 1}{N}$. 

Now, if $\hat{s}= \{\mathcal{a}_k, \mathcal{a}_m, \mathcal{a}_0, \ldots, \mathcal{a}_0\}$, to count the number of elements from $\hat{r}$ to $\hat{s}$, it should be considered subsets of $N-1$ elements, by dropping the first common element $\mathcal{a}_k$. Then, the number of elements from $\hat{r}$ to $\hat{s}$ is the number of elements from $\hat{0}$ to $\{\mathcal{a}_m, \mathcal{a}_0, \ldots, \mathcal{a}_0\}$. They are the combinations of $N-1$ elements taken $\delta_{\mathcal{a}}(\mathcal{a}_m) - 1= m -1$ at a time, with repetitions, i.e., $\binom{\delta_{\mathcal{a}}(\hat{r}_1) + N - 1}{N}$. In general, the positive valuations with constant assignment to the join-irreducible elements can also be expressed as:
\begin{equation}
\label{eq:total-set}
M(\hat{r}) = k \cdot \sum_{j=1}^N \binom{\delta_{\mathcal{a}}(\hat{r}_j) + N - j}{N - j + 1}
\end{equation}

Once, it has been seen that $(R(N), \wedge, \vee, \preceq)$ is a metric lattice and it has been determined all the possible metrics on $R(N)$, it can be explored the metric properties of the IR evaluation measures defined on it. As $(R(N), \preceq )$ is a totally ordered set, every IR evaluation measure is a valuation on this ordering. However, the generalized precision and recall are not isotone on this partial order. For instance, for three relevance degrees ($c=2$), when $N=4$, $\hat{r} = \{\mathcal{a}_1, \mathcal{a}_1, \mathcal{a}_1, \mathcal{a}_0 \} \preceq \{\mathcal{a}_2, \mathcal{a}_0, \mathcal{a}_0, \mathcal{a}_0\} = \hat{s}$. However, $gP(\hat{s}) = 0.25 < 0.375 = gP(\hat{r})$. The same result is valid for the generalized recall since it reaches the same values than the generalized precision. Thus, the generalized precision and recall are not pseudo-metrics on the \textbf{projection+replacement} [set-based] ordering, thus they cannot represent the natural distance. A simple inspection of the analytical expressions of $gP$ and $gR$ it is sufficient to see that they are not similar to Equation \ref{eq:total-set}.

\subsection{Replacement [set-based]}
Any pair of system runs, $\hat{r}, \hat{s} \in R(N)$, such that $\hat{r} \neq \hat{s}$, is ordered by:
$$\hat{r} \preceq \hat{s} \Longleftrightarrow \big\vert\{i : \hat{r}_i \succcurlyeq \mathcal{a}_j\}\big\vert \leq \big\vert\{i : \hat{s}_i \succcurlyeq \mathcal{a}_j\}\big\vert, \forall j \in \{0, \ldots, c\}$$
This approach considers a run greater than another if, fixing any relevance degree, it has more documents above that relevance degree than does the other. In \cite{giner2022effect}, it is shown that this is a partial order and $(R(N), \wedge, \vee, \preceq)$ is a distributive lattice. 

The following result determines the number of join-irreducible elements smaller than a judged run.
\begin{proposition}
\label{prop:join-smaller-set}
Considering $(R(N), \wedge, \vee, \preceq)$, where $\preceq$ is the \textbf{replacement} [set-based] ordering and the indicator function on $REL$, $\delta_{\mathcal{a}}$, then,
$$\big\vert J_{\hat{r}} \big\vert = \sum_{i=1}^N \delta_{\mathcal{a}}(\hat{r}_i)$$
\end{proposition}

$(R(N), \wedge, \vee, \preceq)$ is a metric lattice since it is distributive (see Section \ref{sec:lattice}). In the following, it will be determined all the possible isotone\footnote{They are really positive since $R(N)$ is a finite distributive lattice.} valuations on $R(N)$. Considering an assignment, $w:J_{R(N)} \longrightarrow \mathbb{R}^{+}$ and the Equation \ref{eq:valuation2}; then, any isotone valuation can be expressed as $M(\hat{r})=\sum_{\hat{j} \in J_{\hat{r}}} w(\hat{j})$. Each of these isotone valuations, $M$, generates a metric, $d_M$, which coincides with the minimum path metric and is given by $d_{M}(\hat{r}, \hat{s}) = M(\hat{r}) + M(\hat{s}) - 2 \cdot M(\hat{r} \wedge \hat{s})$, $\forall \hat{r}$, $\hat{s} \in R(N)$.

If $w(\hat{j}) = k$, $\forall \hat{j} \in J_{R(N)}$, with $k \in \mathbb{R}^{+}$, then the natural distance on $(R(N), \preceq)$ is $d_{M}(\hat{r}, \hat{s}) = k \cdot \big\vert J_{\hat{s}} \triangle J_{\hat{r}} \big\vert$ and, by Proposition \ref{prop:join-smaller-set}, the isotone valuations are expressed as:
\begin{equation}
\label{eq:val-set-partial}
M(\hat{r}) = k \cdot \sum_{i=1}^N \delta_{\mathcal{a}}(\hat{r}_i)
\end{equation}

Once, it has been seen that $(R(N), \wedge, \vee, \preceq)$ is a metric lattice and it has been determined all the possible metrics on $R(N)$, it can be explored the metric properties of the IR evaluation measures defined on this partial order. In \cite{giner2022effect}, it was shown that the generalized precision and recall are valuations on the \textbf{replacement} [rank-based] ordering, the following result shows that they are positive.
\begin{proposition}
\label{prop:prec-set-partial}
The generalized precision and recall are positive valuations, when they are defined in $(R(N), \preceq)$ with the \textbf{replacement} [set-based] ordering.
\end{proposition}

As the the generalized precision, $gP$, and recall, $gR$, are positive valuations, then they are metrics on this partial order. In addition, they have a constant assignment to the join-irreducible elements since they can be obtained from Equation \ref{eq:val-set-partial}, when $k= \frac{1}{N \cdot \mathcal{a}_c}$ and $k= \frac{1}{RB \cdot \mathcal{a}_c}$ respectively. Thus, they are the natural distance metric on the \textbf{replacement} [set-based] partial order.

\section{Rank-based Retrieval}
\label{sec:rank-based}
In this Section, the rank-based retrieval is considered and the three orderings which were deduced in \cite{giner2022effect} are analysed.

\subsection{Projection+Replacement [rank-based]}
Let $\hat{r}$, $\hat{s} \in R(N)$ such that $\hat{r} \neq \hat{s}$, then there exists $k = min \{j \leq N : \hat{r}[j] \neq \hat{s}[j] \}$. Any pair of distinct system runs is ordered by: 
$$\hat{r} \preceq \hat{s} \Longleftrightarrow \hat{r}[k] \preccurlyeq \hat{s}[k]$$

A system run is preferred to another if it has a higher relevance degree in the highest rank position at which the two system runs differ. $(R(N), \preceq)$ is a totally ordered set \cite{giner2022effect}.

As $(R(N), \preceq)$ is a totally ordered set \cite{giner2022effect}, then every element is join-irreducible, except $\hat{0}$, i.e., $J_{R(N)} = R(N) \setminus \{\hat{0}\}$. Thus, $J_{\hat{r}}$ is the set of all the elements smaller than $\hat{r}$, except $\hat{0}$.

$(R(N), \wedge, \vee, \preceq)$ is a metric lattice since it is a chain (see Section \ref{sec:lattice}). In the following, it will be determined all the possible isotone\footnote{They are really positive since $R(N)$ is a finite distributive lattice.} valuations on $R(N)$. Consider an assignment, $w:R(N) \setminus \{\hat{0}\} \longrightarrow \mathbb{R}^{+}$ and a judged run $\hat{r} \in R(N) \setminus \{\hat{0}\}$, it can be assumed that $\hat{0} =\hat{r}^{0} \preceq \hat{r}^{1} \preceq \ldots \preceq \hat{r}^{k} = \hat{r}$. As every totally ordered set is a distributive lattice, then, by Equation \ref{eq:valuation2}, any isotone valuation can be obtained as:
$$M(\hat{r}) = \sum_{i = 1}^{k} w(\hat{r}^{i})$$

Considering $\hat{r}$, $\hat{s} \in R(N)$, each of these isotone valuations, $M$, generates a metric, $d_M$, which coincides with the minimum path metric and is given by $d_{M}(\hat{r}, \hat{s}) = M(\hat{r}) + M(\hat{s}) - 2 \cdot M(\hat{r} \wedge \hat{s})$. As $R(N)$, endowed with the \textbf{projection+replacement} [rank-based] ordering, is a totally ordered set; then, it can be assumed that $\hat{r} \preceq \hat{s}$, i.e., $\hat{r} \wedge \hat{s} = \hat{r}$. Therefore, $d_{M}(\hat{r}, \hat{s}) = M(\hat{s}) - M(\hat{r})$.

If $w(\hat{r}) = k$, $\forall \hat{r} \in R(N) \setminus \{\hat{0}\}$, with $k \in \mathbb{R}^{+}$, then these valuations are expressed as $M(\hat{r}) = k \cdot \big\vert J_{\hat{r}} \big\vert$ and the natural distance on the covering graph is $d_{M}(\hat{r}, \hat{s}) = k \cdot \big\vert J_{\hat{s}} \triangle J_{\hat{r}} \big\vert$.

Thus, the isotone valuations, $M(\hat{r})$, with constant assignment to the join-irreducible elements count the number of elements from $\hat{0}$ to $\hat{r}$. It can be given an alternative expression of $M(\hat{r})$. Considering the binary case ($c=1$), the \textbf{projection+replacement} [rank-based] ordering is the following: $(\mathcal{a}_0, \mathcal{a}_0, \mathcal{a}_0, \ldots, \mathcal{a}_0) \precdot (\mathcal{a}_1, \mathcal{a}_0, \mathcal{a}_0, \ldots, \mathcal{a}_0) \precdot (\mathcal{a}_0, \mathcal{a}_1, \mathcal{a}_0, \ldots, \mathcal{a}_0) \precdot (\mathcal{a}_1, \mathcal{a}_1, \mathcal{a}_0, \ldots, \mathcal{a}_0) \precdot (\mathcal{a}_0, \mathcal{a}_0, \mathcal{a}_1, \ldots, \mathcal{a}_0) \precdot (\mathcal{a}_1, \mathcal{a}_0, \mathcal{a}_1, \ldots, \mathcal{a}_0) \precdot (\mathcal{a}_1, \mathcal{a}_1, \mathcal{a}_1, \ldots, \mathcal{a}_0) \precdot \cdots$. That is, they are the sequence of the ordinal numbers expressed in base $2$. In general, the system runs of this ordering can be represented by the conversion in base $10$ of the numbers in base $c+1$ identified by $\delta_{\mathcal{a}}(\hat{r})$, and the ordering among system runs, $\preceq$, corresponds to the ordering, $\leq$, among numbers in base $c+1$. Thus, the positive valuations with constant assignment to the join-irreducible elements can also be expressed as:
\begin{equation}
\label{eq:interval-rank-2}
M(\hat{r}) = k \cdot \sum_{j=1}^N \delta_{\mathcal{a}}(\hat{r}[j]) \cdot (c+1)^{N-j}
\end{equation} 

Once, it has been seen that $(R(N), \wedge, \vee, \preceq)$ is a metric lattice and it has been determined all the possible metrics on $R(N)$, it can be explored the metric properties of the IR evaluation measures defined on this partial order. As $R(N)$ is a totally ordered set, every IR evaluation measure is a valuation on this ordering. However, the generalized precision and recall are not isotone, for instance, in the binary case ($c=1$), for $N=4$, it can be easily checked that $\hat{r} = (\mathcal{a}_0, \mathcal{a}_0, \mathcal{a}_1, \mathcal{a}_1) \preceq (\mathcal{a}_1, \mathcal{a}_0, \mathcal{a}_0, \mathcal{a}_0) = \hat{s}$. However, $gP(\hat{s}) = 0.25 < 0.5 = gP(\hat{r})$. The same result is valid for the generalized recall, since, in the binary case, it has the same values than the generalized precision.

On the other hand, the average precision is not an isotone valuation on the \textbf{projection+replacement} [rank-based] ordering, considering the same example of the previous paragraph, it can be checked that $AP(\hat{s}) = 0.125 < 0.208 = AP(\hat{r})$. The following result shows an interesting property of the $RBP$ evaluation measure.
\begin{proposition}[M. Ferrante \cite{ferrante2018general}]
\label{prop:ferrante}
Considering the ordering \textbf{projection+replacement} [rank-based], $RBP$ is an isotone valuation if and only if $p \leq G/(G + 1)$, where $G= \min_{j \in \{1, \ldots, c\}} (g(\mathcal{a}_j) - g(\mathcal{a}_{j-1})) / g(\mathcal{a}_c)$ is the normalized smallest gap between the gain of two consecutive relevance degrees.
\end{proposition}

Finally, $DCG_b$ is not an isotone valuation with this partial ordering. For instance, in the binary case ($c=1$), when $N=4$; then, $\hat{r} = (\mathcal{a}_0, \mathcal{a}_0, \mathcal{a}_1, \mathcal{a}_1) \preceq (\mathcal{a}_0, \mathcal{a}_1, \mathcal{a}_0, \mathcal{a}_0) = \hat{s}$. However, $DCG_2(\hat{s}) = 1.0 < 1.131 = DCG_2(\hat{r})$.

In all the cases where the IR evaluation measures are not isotone valuations, it follows that they are not pseudo-metrics on the \textbf{projection+replacement} [rank-based] ordering, thus they cannot represent the natural distance. The isotone valuation $RBP$ with the above mentioned restrictions is a pseudo-metric, which is not the natural distance, since it can not be expressed in terms of the Equation \ref{eq:interval-rank-2}.

\subsection{Replacement [rank-based]}
Any pair of system runs, $\hat{r}, \hat{s} \in R(N)$, such that $\hat{r} \neq \hat{s}$, is ordered by:
$$\hat{r} \preceq \hat{s} \Longleftrightarrow \hat{r}[k] \preccurlyeq \hat{s}[k], \ \  \forall k \in \{1, \ldots, N\}$$
A run is greater than another if, for each rank position, it has higher relevance degrees. In \cite{giner2022effect}, it is shown that this is a partial order and $(R(N), \wedge, \vee, \preceq)$ is a distributive lattice. 

The following result determines the number of join-irreducible elements smaller than a judged run.
\begin{proposition}
\label{prop:join-smaller-rank}
Considering $(R(N), \wedge, \vee, \preceq)$, where $\preceq$ is the \textbf{replacement} [rank-based] ordering and the indicator function on $REL$, $\delta_{\mathcal{a}}$, then,
$$\big\vert J_{\hat{r}} \big\vert = \sum_{i=1}^N \delta_{\mathcal{a}}(\hat{r}[i])$$
\end{proposition}

$(R(N), \wedge, \vee, \preceq)$ is a metric lattice since it is distributive (see Section \ref{sec:lattice}). In the following, it will be determined all the possible isotone\footnote{They are really positive since $R(N)$ is a finite distributive lattice.} valuations on $R(N)$. Considering an assignment, $w:J_{R(N)} \longrightarrow \mathbb{R}^{+}$ and the Equation \ref{eq:valuation2}, any positive valuation can be expressed as $M(\hat{r})=\sum_{\hat{j} \in J_{\hat{r}}} w(\hat{j})$. Each of these isotone valuations, $M$, generates a metric, $d_M$, which coincides with the minimum path metric and is given by $d_{M}(\hat{r}, \hat{s}) = M(\hat{r}) + M(\hat{s}) - 2 \cdot M(\hat{r} \wedge \hat{s})$, $\forall \hat{r}$, $\hat{s} \in R(N)$.

If $w(\hat{j}) = k$, $\forall \hat{j} \in J_{R(N)}$, with $k \in \mathbb{R}^{+}$, then the natural distance on $(R(N), \preceq)$ is $d_{M}(\hat{r}, \hat{s}) = k \cdot \big\vert J_{\hat{s}} \triangle J_{\hat{r}} \big\vert$ and, by Proposition \ref{prop:join-smaller-rank}, the isotone valuations are expressed as:
\begin{equation}
\label{eq:val-rank-partial}
M(\hat{r}) = k \cdot \sum_{i=1}^N \delta_{\mathcal{a}}(\hat{r}_i)
\end{equation}

Once, it has been seen that $(R(N), \wedge, \vee, \preceq)$ is a metric lattice and it has been determined all the possible metrics on $R(N)$, it can be explored the metric properties of the IR evaluation measures defined on this partial order. In \cite{giner2022effect}, it was shown that the generalized precision and recall are valuations on the \textbf{replacement} [rank-based] ordering, the following result shows that they are positive.
\begin{proposition}
\label{prop:prec-rank-partial}
The generalized precision and recall are positive valuations, when they are defined in $(R(N), \preceq)$ with the \textbf{replacement} [rank-based] ordering.
\end{proposition}

As the the generalized precision, $gP$, and recall, $gR$, are positive valuations, then they are metrics on this partial order. In addition, they have a constant assignment to the join-irreducible elements since they can be obtained from Equation \ref{eq:val-rank-partial}, when $k= \frac{1}{N \cdot \mathcal{a}_c}$ and $k= \frac{1}{RB \cdot \mathcal{a}_c}$ respectively. Thus, they are the natural distance metric on the \textbf{replacement} [rank-based] partial order.

Similar results are obtained for $RBP$ and $DCG_b$. In \cite{giner2022effect}, it was shown that these IR evaluation measures are valuations on the \textbf{replacement} [rank-based] ordering, the following result shows that they are positive.
\begin{proposition}
\label{prop:rbp-dcg-rank-partial}
$RBP$ and $DCG_b$ are positive valuations, when they are defined in $(R(N), \preceq)$ with the \textbf{Replacement} [rank-based] ordering.
\end{proposition}

As $RBP$ and $DCG_b$ are positive valuations, then they are metrics on this partial order. However, they have not a constant assignment to the join-irreducible elements. Thus, they are not the natural distance metric on the \textbf{replacement} [rank-based] partial order.

\subsection{Swapping+Replacement [rank-based]}
Given two system runs $\hat{r}$, $\hat{s} \in R(N)$, then
\begin{align*}
 \hat{r} &\preceq \hat{s} & &\Longleftrightarrow & \big\vert\{i \leq k : \hat{r}[i] \succcurlyeq \mathcal{a}_j\}\big\vert &\leq \big\vert\{i \leq k : \hat{s}[i] \succcurlyeq \mathcal{a}_j\}\big\vert & \\
& & & & \forall j \in \{0, \ldots, c\}, & \forall k \in \{1, \ldots, N\} &
\end{align*}
A run is considered larger than another one when, for each rank position, it has more relevant documents than the other up to that rank for every relevance degree. 

$(R(N), \wedge, \vee, \preceq)$ is a finite lattice, which contains a $N_5$ sublattice \cite{giner2022effect}, i.e., it is not a modular lattice \cite{birkhoff1940lattice}. By the results of Section \ref{sec:lattice}, $(R(N), \wedge, \vee, \preceq)$ is not a metric lattice. In this case, there cannot exists a positive valuation defined on $R(N)$. As it was noted in \cite{giner2022effect}, there does not exist the natural distance on the covering graph of $R(N)$.

However, there may exist isotone valuations defined on $R(N)$, which are weaker valuations. In these case, the isotone valuation generates a pseudo-metric, which satisfy the same postulates of the metrics, except $d_M(\hat{r},\hat{s})=0 \Leftrightarrow \hat{r}=\hat{s}$.

Once, it has been determined the metric properties of $(R(N), \wedge, \vee, \preceq)$, it can be determined the metric properties of the IR evaluation measures defined on this partial order. The generalized precision is not a positive valuation, for instance, in the binary case for $N=3$, $\hat{r} = (\mathcal{a}_1, \mathcal{a}_0, \mathcal{a}_1) \preceq (\mathcal{a}_1, \mathcal{a}_1, \mathcal{a}_0) = \hat{s}$ and $\hat{r} \neq \hat{s}$, when considering the swapping property at the two last positions. However, $gP(\hat{r}) = 0.22 = gP(\hat{s})$. The same result is valid for the generalized recall since it reaches the same values than the generalized precision.

However, the following results show that the IR evaluation measures are isotone valuations on $(R(N), \wedge, \vee, \preceq)$.

\begin{proposition}
\label{prop:prec-rank-swap}
The generalized precision and recall are isotone valuations, when they are defined on $(R(N), \wedge, \vee, \preceq)$ with the \textbf{Swapping+Replacement} [rank-based] ordering.
\end{proposition}

\begin{proposition}
\label{prop:rbp-dcg-rank-swap}
$RBP$ and $DCG_b$ are isotone valuations, when they are defined on $(R(N), \wedge, \vee, \preceq)$ with the \textbf{Swapping+Replacement} [rank-based] ordering.
\end{proposition}

These results indicate that $(R(N), \wedge, \vee, \preceq)$ is a pseudo-metric lattice, where the generalized precision and recall, $RBP$ or $DCG_b$ are pseudo-metrics.

\section{Related Work}
\label{sec:SoA}
The formal analysis of IR evaluation metrics has contributed to a better understanding of their properties. One of the first attempts is given in \cite{swets1963information}, where the effectiveness and the efficiency of IR evaluation measures are analysed in terms of a $2$-by-$2$ contingency table of pertinence and retrieval. Later, van Rijsbergen \cite{van1974foundation,van1979information}, tackle the issue of the foundations of measurement in IR through a conjoint (additive) structure based on precision and recall; then, he examines the properties of a measure on this prec-recall structure. In \cite{bollmann1980measurement}, a similar conjoint structure is defined, but on the contingency table of the binary retrieval; then, they study the properties of the proposed MZ-metric. In \cite{yao1995measuring}, user judgements on documents are formally described by a weak order, then, a measure of system performance is presented, whose appropriateness is demonstrated through an axiomatic approach. In \cite{huibers1996axiomatic}, a framework for the theoretical comparison of IR models, based on situation theory, is presented. It allows an inference mechanism with the axiomatised concept of \emph{aboutness}. In \cite{busin2013axiometrics, maddalena2014axiometrics} an axiomatic definition of IR effectiveness metric is provided, in a unifying framework for ranking, classification and clustering. In \cite{sebastiani2020evaluation}, it is discussed what properties should enjoy an evaluation measure for quantification. 

In the same formal research line, some works have analysed desirable properties that any suitable IR evaluation measure should satisfy. The three metric postulates, studied in this paper, are implicitly considered in some of these properties or axioms. In \cite{sebastiani2015axiomatically}, the axioms that an evaluation measure for classification should satisfy are discussed, then the K evaluation metric is proposed. In  \cite{moffat2013seven}, the evaluation measures are characterized by seven numeric properties, such as \emph{convergence}, \emph{top-weightedness} and \emph{localization}, which provides a framework to classify the measures according to their effectiveness. In \cite{amigo2009comparison, amigo2013general, amigo2018axiomatic}, the properties of \emph{priority}, \emph{deepness}, \emph{closeness threshold}, \emph{confidence}, etc., are stated to derive evaluation measures, which can be applied to several information access tasks. Recently, in \cite{giner2022effect}, three of these ranking axioms are considered; then, the possible orderings in the set of rankings of documents are derived. 

One of the closest works to the present is \cite{bollmann1984two}, where two axioms are considered: the \emph{monotonicity}, also known as the \emph{homogeneity law} in measurement theory and the \emph{Archimedian}, then they determine explicit expressions of the evaluation measures in the binary retrieval, by supporting the results on measurement theory. However, this paper considers ranking axioms, which are directly related to preferences on retrieval measures. In addition, it is grounded on lattice theory and the results are generalized to multi-grade relevance. The other close work is \cite{ferrante2018general}, where a theory of IR evaluation measures, based on the representational theory of measurements, is developed to determine whether and when IR measures are interval scales. Some orderings on the set of rankings are derived from two operations: replacement and swap; then, the results are proven by a common generic procedure, using the notion of interval and differences. However, this work is based on ranking orderings derived from three operations, which introduces a new ordering. In addition, the concepts of metric and pseudo-metric are less restrictive than the concepts of ordinal and interval scales, thus the results of this paper generalize some results of \cite{ferrante2018general}.

\section{Conclusions}
\label{sec:conclusion}
The properties of IR evaluation measures do not depend only on its analytical expression. The empirical domain where they are defined play an important role. This set can be characterized by desirable properties that any reasonable IR evaluation measure should satisfy. These properties lead to different orderings in the set of ranked lists of documents \cite{giner2022effect}. This paper checks if the set of rankings endowed with these orderings are metric or pseudo-metric spaces and determine the possible metrics and pseudo-metrics defined on it. There are shown explicit expressions of these mappings, in terms of a remarkable subset of system runs: the join-irreducible elements. It is found that, when the relevant documents are prioritised, precision, recall, $RBP$ and $DCG$ are metrics on the set of rankings, however they are pseudo-metrics when the swapping of documents is considered. Thus, it has no sense to state that ``a specific IR evaluation measure is a metric or pseudo-metric'', since it has to be considered the empirical domain where it is defined.

The use of evaluation measures is a common practice to evaluate, benchmark and track the effectiveness of IR systems. Then, a natural question arises: which measure should be chosen? A (partial) answer to this question is given by the formal analysis of their metric properties, more specifically, by their axiomatic characterization. Some works have characterised when IR evaluation measures are ordinal scales ($a \geq b \Leftrightarrow f(a) \geq f(b)$) \cite{ferrante2015towards} or interval scales \cite{ferrante2018general}, in the orderings of \cite{giner2022effect}. Here, these results have been generalised, by determining the pseudo-metrics, or equivalently, the valuations that verify: $a \geq b \Rightarrow f(a) \geq f(b)$. 

\appendix
\section*{Appendix}

{\bf Proposition \ref{prop:join-smaller-set}:} \textit{Considering $(R(N), \wedge, \vee, \preceq)$, where $\preceq$ is the \textbf{replacement} [set-based] ordering and the indicator function on $REL$, $\delta_{\mathcal{a}}$, then,}
\[ \big\vert J_{\hat{r}} \big\vert = \sum_{i=1}^N \delta_{\mathcal{a}}(\hat{r}_i) \]
\begin{proof}
Consider the convention to represent $\hat{r}$ as noted in Section \ref{sec:notation}; then, it can be assumed that $\hat{r}=\{\mathcal{a}_h, \mathcal{a}_j, \mathcal{a}_k, \ldots\}$, where $0 \leq \cdots \leq k \leq j \leq h \leq c$. In \cite{giner2022effect}, it is shown that the joint-irreducible elements of $R(N)$ with the \textbf{replacement} [set-based] partial ordering, are:
\begin{small}
\begin{equation} \label{eq:join-irred-set}
\begin{matrix}
\{\mathcal{a}_1, \mathcal{a}_0, \ldots, \mathcal{a}_0 \}, & \{\mathcal{a}_1, \mathcal{a}_1, \mathcal{a}_0, \ldots, \mathcal{a}_0 \}, & \ldots, & \{\mathcal{a}_1, \ldots, \mathcal{a}_1 \},  \\
\{\mathcal{a}_2, \mathcal{a}_0, \ldots, \mathcal{a}_0 \}, & \{\mathcal{a}_2, \mathcal{a}_2, \mathcal{a}_0, \ldots, \mathcal{a}_0 \}, & \ldots, & \{\mathcal{a}_2, \ldots, \mathcal{a}_2 \},  \\
\vdots & \vdots & \vdots & \vdots  \\
\{\mathcal{a}_c, \mathcal{a}_0, \ldots, \mathcal{a}_0 \}, & \{\mathcal{a}_c, \mathcal{a}_c, \mathcal{a}_0, \ldots, \mathcal{a}_0 \}, & \ldots, & \{\mathcal{a}_c, \ldots, \mathcal{a}_c \} 
\end{matrix}
\end{equation}
\end{small} 

On the other hand, the subset of join-irreducible elements smaller that $\hat{r}$, i.e., $J_{\hat{r}}$, can be seen as the union of the following subsets of join-irreducible elements: (i) there are $h = \delta_{\mathcal{a}}(\hat{r}_1)$ join-irreducible elements corresponding to the first judged document, $\mathcal{a}_h$, of $\hat{r}$: $\{\mathcal{a}_1, \mathcal{a}_0, \ldots, \mathcal{a}_0 \},  \{\mathcal{a}_2, \mathcal{a}_0, \ldots, \mathcal{a}_0 \},  \ldots, \{\mathcal{a}_h, \mathcal{a}_0, \ldots, \mathcal{a}_0 \}$. Note that by definition of the partial order, $\preceq$, these elements are smaller than $\hat{r}$; (ii) there are $j = \delta_{\mathcal{a}}(\hat{r}_2)$ join-irreducible elements corresponding to the second judged document, $\mathcal{a}_j$, of $\hat{r}$: $\{\mathcal{a}_1, \mathcal{a}_1, \mathcal{a}_0, \ldots, \mathcal{a}_0 \},  \{\mathcal{a}_2, \mathcal{a}_2, \mathcal{a}_0, \ldots, \mathcal{a}_0 \},  \ldots, \{\mathcal{a}_j, \mathcal{a}_j, \mathcal{a}_0, \ldots, \mathcal{a}_0 \}$. Note that by definition of the partial order, $\preceq$, these elements are smaller than $\hat{r}$; (iii) and so on.

Now, suppose that there exists a join-irreducible element smaller than $\hat{r}$ and different from the elements listed in the previous paragraph; then, inspecting the form of the possible join-irreducible elements in Equation \ref{eq:join-irred-set}, there exists a join-irreducible, $\hat{m} = \{\mathcal{a}_m, \ldots, \mathcal{a}_m, \mathcal{a}_0, \ldots, \mathcal{a}_0\}$, such that $\hat{m} \preceq \hat{r}$ and $p$ is the largest position of $\hat{m}$ where the judged document is classified as $\mathcal{a}_m$.

It can be assumed that $\hat{r} = \{\mathcal{a}_h, \mathcal{a}_j, \mathcal{a}_k, \ldots, \mathcal{a}_n, \ldots\}$, where $\mathcal{a}_n$ is placed at position $p$, $\mathcal{a}_n \preccurlyeq \mathcal{a}_m$ and $\mathcal{a}_n \neq \mathcal{a}_m$. By Theorem \ref{th:join-join-irreducible}, in a finite distributive lattice, every element can be expressed in a unique manner as a join of the join-irreducible elements. Note that these join-irreducible elements are smaller than the element since it is join of them. As $(R(N), \wedge, \vee, \preceq)$ is a distributive lattice with the \textbf{replacement} [set-based] partial order, thus, $\hat{r} = \bigvee J_{\hat{r}} \vee \hat{m}$. In addition, in \cite{giner2022effect}, it is shown that the join operation in the \textbf{replacement} [set-based] partial order is the maximum component wise. However, this contradicts the last equality since the judged document of $\hat{r}$ at position $p$ is $\mathcal{a}_n$ and the judged document of $\bigvee J_{\hat{r}} \vee \hat{m}$ at position $p$ is $\mathcal{a}_m$.

Therefore, the elements listed bellow, as subsets of $J_{\hat{r}}$, are the join-irreducible elements smaller than $\hat{r}$ and it holds:
\[ \big\vert J_{\hat{r}} \big\vert = \sum_{i=1}^N \delta_{\mathcal{a}}(\hat{r}_i) \]
\end{proof}

{\bf Proposition \ref{prop:prec-set-partial}:} \textit{The generalized precision and recall are positive valuations, when they are defined in $(R(N), \preceq)$ with the \textbf{replacement} [set-based] ordering.}
\begin{proof}
The generalized precision will be a positive valuation if the implication: $\hat{r} \preceq \hat{s}$ and $\hat{r} \neq \hat{s} \Rightarrow gP(\hat{r}) < gP(\hat{s})$ holds true. 

In \cite{giner2022effect}, it is shown that the covering relation of the \textbf{replacement} [set-based] ordering is as follows: given a pair of system runs, $\hat{r}, \hat{s} \in R(N)$ with $\hat{r} \neq \hat{s}$, then $\hat{r} \precdot \hat{s}$ $\Leftrightarrow$ $\exists k : \hat{r}_k = \mathcal{a}_i, \hat{s}_k = \mathcal{a}_{i+1}$ and $\hat{r}_j = \hat{s}_j, \forall j \neq k$.

Thus, if $\hat{r} \preceq \hat{s}$ and $\hat{r} \neq \hat{s}$, then there exists at least $k$, such that $\hat{r}_k \prec \hat{s}_k$ and $\hat{r}_j \preccurlyeq \hat{s}_j, \forall j \neq k$. Considering that $g$ is an strictly increasing function, it follows: 
$$\hat{r} \preceq \hat{s},\ \ \hat{r} \neq \hat{s}  \Longrightarrow \sum_{i=1}^N g(\hat{r}_i) < \sum_{i=1}^N g(\hat{s}_i) \ .$$

As $N$ and $g(\mathcal{a}_c)$ are constants,
$$gP(\hat{r}) = \frac{1}{N} \cdot \sum_{i=1}^{N} \frac{g(\hat{r}_i)}{g(\mathcal{a}_{c})} < \frac{1}{N} \cdot \sum_{i=1}^{N} \frac{g(\hat{s}_i)}{g(\mathcal{a}_{c})} = gP(\hat{s}) \ .$$
Similarly, as $RB$ is constant,
$$gR(\hat{r}) = \frac{1}{RB} \cdot \sum_{i=1}^{N} \frac{g(\hat{r}_i)}{g(\mathcal{a}_{c})} < \frac{1}{RB} \cdot \sum_{i=1}^{N} \frac{g(\hat{s}_i)}{g(\mathcal{a}_{c})} = gR(\hat{s})$$
\end{proof}

{\bf Proposition \ref{prop:join-smaller-rank}:} \textit{Considering $(R(N), \wedge, \vee, \preceq)$, where $\preceq$ is the \textbf{replacement} [rank-based] ordering and the indicator function on $REL$, $\delta_{\mathcal{a}}$, then,}
\[ \big\vert J_{\hat{r}} \big\vert = \sum_{i=1}^N \delta_{\mathcal{a}}(\hat{r}[i]) \]
\begin{proof}
It can be assumed that $\hat{r}=(\mathcal{a}_h, \mathcal{a}_j, \mathcal{a}_k, \ldots)$, where $\mathcal{a}_h$, $\mathcal{a}_j$, $\mathcal{a}_k$, $\ldots \in REL$. In \cite{giner2022effect}, it is shown that the joint-irreducible elements of $R(N)$ with the \textbf{replacement} [rank-based] partial ordering, are:
\begin{small}
\begin{equation} \label{eq:join-irred-rank}
\begin{matrix}
(\mathcal{a}_1, \mathcal{a}_0, \ldots, \mathcal{a}_0 ), & (\mathcal{a}_0, \mathcal{a}_1, \mathcal{a}_0, \ldots, \mathcal{a}_0 ), & \ldots, & (\mathcal{a}_0, \mathcal{a}_0, \ldots, \mathcal{a}_1),  \\
(\mathcal{a}_2, \mathcal{a}_0, \ldots, \mathcal{a}_0 ), & (\mathcal{a}_0, \mathcal{a}_2, \mathcal{a}_0, \ldots, \mathcal{a}_0 ), & \ldots, & (\mathcal{a}_0, \mathcal{a}_0, \ldots, \mathcal{a}_2 ),  \\
\vdots & \vdots & \vdots & \vdots \\
(\mathcal{a}_c, \mathcal{a}_0, \ldots, \mathcal{a}_0 ), & (\mathcal{a}_0, \mathcal{a}_c, \mathcal{a}_0, \ldots, \mathcal{a}_0), & \ldots, & (\mathcal{a}_0, \mathcal{a}_0, \ldots, \mathcal{a}_c) 
\end{matrix}
\end{equation}
\end{small}

On the other hand, the subset of join-irreducible elements smaller that $\hat{r}$, i.e., $J_{\hat{r}}$, can be seen as the union of the following subsets of join-irreducible elements: (i) there are $h = \delta_{\mathcal{a}}(\hat{r}[1])$ join-irreducible elements corresponding to the first judged document, $\mathcal{a}_h$, of $\hat{r}$: $(\mathcal{a}_1, \mathcal{a}_0, \ldots, \mathcal{a}_0 ),  (\mathcal{a}_2, \mathcal{a}_0, \ldots, \mathcal{a}_0),  \ldots, (\mathcal{a}_h, \mathcal{a}_0, \ldots, \mathcal{a}_0)$. Note that by definition of the partial order, $\preceq$, these elements are smaller than $\hat{r}$; (ii) there are $j = \delta_{\mathcal{a}}(\hat{r}[2])$ join-irreducible elements corresponding to the second judged document, $\mathcal{a}_j$, of $\hat{r}$: $(\mathcal{a}_0, \mathcal{a}_1, \mathcal{a}_0, \ldots, \mathcal{a}_0),  (\mathcal{a}_0, \mathcal{a}_2, \mathcal{a}_0, \ldots, \mathcal{a}_0),  \ldots, (\mathcal{a}_0, \mathcal{a}_j, \mathcal{a}_0, \ldots, \mathcal{a}_0)$. Note that by definition of the partial order, $\preceq$, these elements are smaller than $\hat{r}$; (iii) and so on.

Now, suppose that there exists a join-irreducible element smaller than $\hat{r}$ and different from the elements listed in the previous paragraph; then, inspecting the form of the possible join-irreducible elements in Equation \ref{eq:join-irred-rank}, there exists a join-irreducible, $\hat{m} = (\mathcal{a}_0, \ldots,\mathcal{a}_0, \mathcal{a}_m, \mathcal{a}_0, \ldots, \mathcal{a}_0)$, such that $\hat{m} \preceq \hat{r}$ and $p$ is the position of $\mathcal{a}_m$.

It can be assumed that $\hat{r} = (\mathcal{a}_h, \mathcal{a}_j, \mathcal{a}_k, \ldots, \mathcal{a}_n, \ldots)$, where $\mathcal{a}_n$ is placed at position $p$, $\mathcal{a}_n \preccurlyeq \mathcal{a}_m$ and $\mathcal{a}_n \neq \mathcal{a}_m$. By Theorem \ref{th:join-join-irreducible}, in a finite distributive lattice, every element can be expressed in a unique manner as a join of the join-irreducible elements. Note that these join-irreducible elements are smaller than the element since it is join of them. As $(R(N), \wedge, \vee, \preceq)$ is a distributive lattice with the \textbf{replacement} [rank-based] partial order, thus, $\hat{r} = \bigvee J_{\hat{r}} \vee \hat{m}$. In addition, in \cite{giner2022effect}, it is shown that the join operation in the \textbf{replacement} [rank-based] partial order is the maximum coordinate wise. However, this contradicts the last equality since the judged document of $\hat{r}$ at position $p$ is $\mathcal{a}_n$ and the judged document of $\bigvee J_{\hat{r}} \vee \hat{m}$ at position $p$ is $\mathcal{a}_m$.

Therefore, the elements listed bellow, as subsets of $J_{\hat{r}}$, are the join-irreducible elements smaller than $\hat{r}$ and it holds:
\[ \big\vert J_{\hat{r}} \big\vert = \sum_{i=1}^N \delta_{\mathcal{a}}(\hat{r}_i) \]
\end{proof}

{\bf Proposition \ref{prop:prec-rank-partial}:} \textit{The generalized precision and recall are positive valuations, when they are defined in $(R(N), \preceq)$ with the \textbf{replacement} [rank-based] ordering.}
\begin{proof}
The generalized precision will be a positive valuation if the implication: $\hat{r} \preceq \hat{s}$ and $\hat{r} \neq \hat{s} \Rightarrow gP(\hat{r}) < gP(\hat{s})$ holds true. 

In \cite{giner2022effect}, it is shown that the covering relation of the \textbf{replacement} [rank-based] ordering is as follows: given a pair of system runs, $\hat{r}, \hat{s} \in R(N)$ with $\hat{r} \neq \hat{s}$, then $\hat{r} \precdot \hat{s} \Leftrightarrow \exists k : \hat{r}[k] = \mathcal{a}_i, \hat{s}[k] = \mathcal{a}_{i+1}$ and $\hat{r}[j] = \hat{s}[j], \forall j \neq k$.

Thus, if $\hat{r} \preceq \hat{s}$ and $\hat{r} \neq \hat{s}$, then there exists at least $k$, such that $\hat{r}[k] \prec \hat{s}[k]$ and $\hat{r}[j] \preccurlyeq \hat{s}[j], \forall j \neq k$. Considering that $g$ is an strictly increasing function, it follows: 
$$\hat{r} \preceq \hat{s},\ \ \hat{r} \neq \hat{s}  \Longrightarrow \sum_{i=1}^N g(\hat{r}[i]) < \sum_{i=1}^N g(\hat{s}[i]) \ .$$

As $N$ and $g(\mathcal{a}_c)$ are constants,
$$gP(\hat{r}) = \frac{1}{N} \cdot \sum_{i=1}^{N} \frac{g(\hat{r}[i])}{g(\mathcal{a}_{c})} < \frac{1}{N} \cdot \sum_{i=1}^{N} \frac{g(\hat{s}[i])}{g(\mathcal{a}_{c})} = gP(\hat{s}) \ .$$
Similarly, since $RB$ is constant,
$$gR(\hat{r}) = \frac{1}{RB} \cdot \sum_{i=1}^{N} \frac{g(\hat{r}[i])}{g(\mathcal{a}_{c})} < \frac{1}{RB} \cdot \sum_{i=1}^{N} \frac{g(\hat{s}[i])}{g(\mathcal{a}_{c})} = gR(\hat{s})$$
\end{proof}

{\bf Proposition \ref{prop:rbp-dcg-rank-partial}:} \textit{$RBP$ and $DCG_b$ are positive valuations, when they are defined in $(R(N), \preceq)$ with the \textbf{Replacement} [rank-based] ordering.}
\begin{proof}
$RBP$ will be a positive valuation if the implication: $\hat{r} \preceq \hat{s}$ and $\hat{r} \neq \hat{s} \Rightarrow gP(\hat{r}) < gP(\hat{s})$ holds true. 

In \cite{giner2022effect}, it is shown that the covering relation of the \textbf{replacement} [rank-based] ordering is as follows: given a pair of system runs, $\hat{r}, \hat{s} \in R(N)$ with $\hat{r} \neq \hat{s}$, then $\hat{r} \precdot \hat{s} \Leftrightarrow \exists k : \hat{r}[k] = \mathcal{a}_i, \hat{s}[k] = \mathcal{a}_{i+1}$ and $\hat{r}[j] = \hat{s}[j], \forall j \neq k$.

Thus, if $\hat{r} \preceq \hat{s}$ and $\hat{r} \neq \hat{s}$, then there exists at least $k$, such that $\hat{r}[k] \prec \hat{s}[k]$ and $\hat{r}[j] \preccurlyeq \hat{s}[j], \forall j \neq k$. Considering that $g$ is an strictly increasing function, it follows: 
$$\hat{r} \preceq \hat{s},\ \ \hat{r} \neq \hat{s}  \Longrightarrow \sum_{i=1}^N g(\hat{r}[i]) < \sum_{i=1}^N g(\hat{s}[i]) \ .$$

Note that if $g(\hat{r}[i]) < g(\hat{s}[i])$, then $p^{i-1} \cdot g(\hat{r}[i]) < p^{i-1} \cdot g(\hat{s}[i])$, $\forall i \in \{1, \ldots, N\}$, then, 
$$gRBP(\hat{r}) = \frac{(1-p)}{g(\mathcal{a}_c)} \cdot \sum_{i=1}^N p^{i-1} \cdot g(\hat{r}[i]) < \frac{(1-p)}{g(\mathcal{a}_c)} \cdot \sum_{i=1}^N p^{i-1} \cdot g(\hat{s}[i]) = gRBP(\hat{s}) \ .$$
And similarly, if $g(\hat{r}[i]) < g(\hat{s}[i])$, then $\frac{g(\hat{r[i]})}{\max \{1, \log_{b}i \}} < \frac{g(\hat{s[i]})}{\max \{1, \log_{b}i \}}$, $\forall i \in \{1, \ldots, N\}$, then
$$DCG_{b}(\hat{r}) = \sum_{i=1}^N \frac{g(\hat{r[i]})}{\max \{1, \log_{b}i \}} < \sum_{i=1}^N \frac{g(\hat{s[i]})}{\max \{1, \log_{b}i \}} = DCG_{b}(\hat{s})$$
\end{proof}

{\bf Proposition \ref{prop:prec-rank-swap}:} \textit{The generalized precision and recall are isotone valuations, when they are defined on $(R(N), \wedge, \vee, \preceq)$ with the \textbf{Swapping+Replacement} [rank-based] ordering.}
\begin{proof}
The generalized precision will be an isotone valuation if the implication: $\hat{r} \preceq \hat{s} \Rightarrow gP(\hat{r}) \leq gP(\hat{s})$ holds true. 

Let $\hat{r}, \hat{s} \in R(N)$ such that $\hat{r} \neq \hat{s}$, by definition of the \textbf{Swapping+Replacement} [rank-based] ordering and considering that $g$ is an strictly increasing function:
\begin{align*}
\hat{r} &\preceq \hat{s} & &\Longleftrightarrow & \big\vert\{i \leq k : \hat{r}[i] \succcurlyeq \mathcal{a}_j\}\big\vert &\leq \big\vert\{i \leq k : \hat{s}[i] \succcurlyeq \mathcal{a}_j\}\big\vert & &\Longleftrightarrow & \sum_{i=1}^k g(\hat{r}[i]) &\leq \sum_{i=1}^k g(\hat{s}[i]),  \\
& & & & \forall j \in \{0, \ldots, c\}, & \forall k \in \{1, \ldots, N\} & & &  \forall \ k \in & \{1, \ldots, N\} 
\end{align*}

In particular, when $k=N$, the last expression is $\sum_{i=1}^N g(\hat{r}[i]) \leq \sum_{i=1}^N g(\hat{s}[i])$.

Considering that $g(\mathcal{a}_c)$ and $N$ are constants,
$$gP(\hat{r}) = \frac{1}{N} \cdot \sum_{i=1}^{N} \frac{g(\hat{r}[i])}{g(\mathcal{a}_{c})} \leq \frac{1}{N} \cdot \sum_{i=1}^{N} \frac{g(\hat{s}[i])}{g(\mathcal{a}_{c})} = gP(\hat{s}) \ .$$
Similarly, as $RB$ is constant,
$$gR(\hat{r}) = \frac{1}{RB} \cdot \sum_{i=1}^{N} \frac{g(\hat{r}[i])}{g(\mathcal{a}_{c})} \leq \frac{1}{RB} \cdot \sum_{i=1}^{N} \frac{g(\hat{s}[i])}{g(\mathcal{a}_{c})} = gR(\hat{s})$$
\end{proof}

{\bf Proposition \ref{prop:rbp-dcg-rank-swap}:} \textit{$RBP$ and $DCG_b$ are isotone valuations, when they are defined on $(R(N), \wedge, \vee, \preceq)$ with the \textbf{Swapping+Replacement} [rank-based] ordering.}
\begin{proof}
$RBP$ will be an isotone valuation if the implication: $\hat{r} \preceq \hat{s} \Rightarrow gRBP(\hat{r}) \leq gRBP(\hat{s})$ holds true. 

Let $\hat{r}, \hat{s} \in R(N)$ such that $\hat{r} \neq \hat{s}$, by definition of the \textbf{Swapping+Replacement} [rank-based] ordering and considering that $g$ is an strictly increasing function: 
\begin{align*}
\hat{r} &\preceq \hat{s} & &\Longleftrightarrow & \big\vert\{i \leq k : \hat{r}[i] \succcurlyeq \mathcal{a}_j\}\big\vert &\leq \big\vert\{i \leq k : \hat{s}[i] \succcurlyeq \mathcal{a}_j\}\big\vert & &\Longleftrightarrow & \sum_{i=1}^k g(\hat{r}[i]) &\leq \sum_{i=1}^k g(\hat{s}[i]),  \\
& & & & \forall j \in \{0, \ldots, c\}, & \forall k \in \{1, \ldots, N\} & & &  \forall \ k \in & \{1, \ldots, N\} 
\end{align*}

In particular, when $k=N$, the last expression is $\sum_{i=1}^N g(\hat{r}[i]) \leq \sum_{i=1}^N g(\hat{s}[i])$.

Notice that if $g(\hat{r}[i]) \leq g(\hat{s}[i])$, then $p^{i-1} \cdot g(\hat{r}[i]) \leq p^{i-1} \cdot g(\hat{s}[i])$, $\forall i \in \{1, \ldots, N\}$, thus
$$gRBP(\hat{r}) = \frac{(1-p)}{g(\mathcal{a}_c)} \cdot \sum_{i=1}^N p^{i-1} \cdot g(\hat{r}[i]) \leq \frac{(1-p)}{g(\mathcal{a}_c)} \cdot \sum_{i=1}^N p^{i-1} \cdot g(\hat{s}[i]) = gRBP(\hat{s}) \ .$$
And similarly, if $g(\hat{r}[i]) \leq g(\hat{s}[i])$, then $\frac{g(\hat{r[i]})}{\max \{1, \log_{b}i \}} \leq \frac{g(\hat{s[i]})}{\max \{1, \log_{b}i \}}$, $\forall i \in \{1, \ldots, N\}$, thus
$$DCG_{b}(\hat{r}) = \sum_{i=1}^N \frac{g(\hat{r[i]})}{\max \{1, \log_{b}i \}} \leq \sum_{i=1}^N \frac{g(\hat{s[i]})}{\max \{1, \log_{b}i \}} = DCG_{b}(\hat{s})$$
\end{proof}

\bibliographystyle{unsrt}  
\bibliography{references}

\end{document}